\documentclass[10pt,conference,nofonttune]{IEEEtran}
\IEEEoverridecommandlockouts
\usepackage[T1]{fontenc}
\usepackage[utf8]{inputenc}
\usepackage[varqu]{zi4}

\usepackage{cite}
\usepackage{amsmath,amssymb,amsfonts}
\usepackage{newtxtext,newtxmath}
\usepackage{graphicx}
\usepackage[footnotesize]{caption}
\usepackage{subcaption}
\usepackage{tikz}
\usetikzlibrary{arrows.meta,patterns,patterns.meta}
\usepackage{pgfplots}
\pgfplotsset{compat=1.18}
\usepgfplotslibrary{groupplots,colormaps}
\usepackage{tabularx,booktabs,array,multirow}
\usepackage{textcomp,xcolor,xspace,soul,ragged2e}
\usepackage{algorithmic,circledsteps,listings}
\usepackage{url}
\usepackage[draft]{hyperref}
\newlength\fheight
\newlength\fwidth
\usepackage[acronyms,nonumberlist,nopostdot,nomain,
            nogroupskip,acronymlists={hidden}]{glossaries}
\newglossary[algh]{hidden}{acrh}{acnh}{Hidden Acronyms}
\glsdisablehyper

\newacronym{3gpp}{3GPP}{3rd Generation Partnership Project}
\newacronym{4g}{4G}{4th generation}
\newacronym{5g}{5G}{5th generation}
\newacronym{6g}{6G}{6th generation}
\newacronym{5gc}{5GC}{5G Core}
\newacronym{adc}{ADC}{Analog to Digital Converter}
\newacronym{aerpaw}{AERPAW}{Aerial Experimentation and Research Platform for Advanced Wireless}
\newacronym{ai}{AI}{Artifial Intelligence}
\newacronym{aimd}{AIMD}{Additive Increase Multiplicative Decrease}
\newacronym{am}{AM}{Acknowledged Mode}
\newacronym{amc}{AMC}{Adaptive Modulation and Coding}
\newacronym{amf}{AMF}{Access and Mobility Management Function}
\newacronym{aops}{AOPS}{Adaptive Order Prediction Scheduling}
\newacronym{api}{API}{Application Programming Interface}
\newacronym{apn}{APN}{Access Point Name}
\newacronym{ap}{AP}{Application Protocol}
\newacronym{aqm}{AQM}{Active Queue Management}
\newacronym{ausf}{AUSF}{Authentication Server Function}
\newacronym{avc}{AVC}{Advanced Video Coding}
\newacronym{awgn}{AGWN}{Additive White Gaussian Noise}
\newacronym{balia}{BALIA}{Balanced Link Adaptation Algorithm}
\newacronym{bbu}{BBU}{Base Band Unit}
\newacronym{bdp}{BDP}{Bandwidth-Delay Product}
\newacronym{ber}{BER}{Bit Error Rate}
\newacronym{bf}{BF}{Beamforming}
\newacronym{bler}{BLER}{Block Error Rate}
\newacronym{brr}{BRR}{Bayesian Ridge Regressor}
\newacronym{bs}{BS}{Base Station}
\newacronym{bsr}{BSR}{Buffer Status Report}
\newacronym{bss}{BSS}{Business Support System}
\newacronym{ca}{CA}{Carrier Aggregation}
\newacronym{caas}{CaaS}{Connectivity-as-a-Service}
\newacronym{cb}{CB}{Code Block}
\newacronym{cc}{CC}{Congestion Control}
\newacronym{ccid}{CCID}{Congestion Control ID}
\newacronym{cco}{CC}{Carrier Component}
\newacronym{cdd}{CDD}{Cyclic Delay Diversity}
\newacronym{cdf}{CDF}{Cumulative Distribution Function}
\newacronym{cdn}{CDN}{Content Distribution Network}
\newacronym{ci}{CI}{Continuous Integration}
\newacronym{cd}{CD}{Continuous Deployment}
\newacronym{ct}{CT}{Continuous Testing}
\newacronym{cli}{CLI}{Command-line Interface}
\newacronym{cn}{CN}{Core Network}
\newacronym{codel}{CoDel}{Controlled Delay Management}
\newacronym{mae}{MAE}{Mean Absolute Error}
\newacronym{comac}{COMAC}{Converged Multi-Access and Core}
\newacronym{cord}{CORD}{Central Office Re-architected as a Datacenter}
\newacronym{cornet}{CORNET}{COgnitive Radio NETwork}
\newacronym{cosmos}{COSMOS}{Cloud Enhanced Open Software Defined Mobile Wireless Testbed for City-Scale Deployment}
\newacronym{cots}{COTS}{Commercial Off-the-Shelf}
\newacronym{cp}{CP}{Control Plane}
\newacronym{cpt}{CPT}{Conditional Probability Table}
\newacronym{cyp}{CP}{Cyclic Prefix}
\newacronym{up}{UP}{User Plane}
\newacronym{cpu}{CPU}{Central Processing Unit}
\newacronym{cqi}{CQI}{Channel Quality Information}
\newacronym{cr}{CR}{Cognitive Radio}
\newacronym{cran}{CRAN}{Cloud \gls{ran}}
\newacronym{crs}{CRS}{Cell Reference Signal}
\newacronym{csi}{CSI}{Channel State Information}
\newacronym{csirs}{CSI-RS}{Channel State Information - Reference Signal}
\newacronym{cu}{CU}{Central Unit}
\newacronym{d2tcp}{D$^2$TCP}{Deadline-aware Data center TCP}
\newacronym{d3}{D$^3$}{Deadline-Driven Delivery}
\newacronym{dac}{DAC}{Digital to Analog Converter}
\newacronym{dag}{DAG}{Directed Acyclic Graph}
\newacronym{das}{DAS}{Distributed Antenna System}
\newacronym{dash}{DASH}{Dynamic Adaptive Streaming over HTTP}
\newacronym{dc}{DC}{Dual Connectivity}
\newacronym{dccp}{DCCP}{Datagram Congestion Control Protocol}
\newacronym{dce}{DCE}{Direct Code Execution}
\newacronym{dci}{DCI}{Downlink Control Information}
\newacronym{codec}{CODEC}{Conditional Dependence Coefficient}
\newacronym{smote}{SMOTE}{Synthetic Minority Over-sampling Technique}
\newacronym{dctcp}{DCTCP}{Data Center TCP}
\newacronym{dl}{DL}{Downlink}
\newacronym{dmr}{DMR}{Deadline Miss Ratio}
\newacronym{dmrs}{DMRS}{DeModulation Reference Signal}
\newacronym{drlcc}{DRL-CC}{Deep Reinforcement Learning Congestion Control}
\newacronym{drs}{DRS}{Discovery Reference Signal}
\newacronym{du}{DU}{Distributed Unit}
\newacronym{e2e}{E2E}{end-to-end}
\newacronym{earfcn}{EARFCN}{E-UTRA Absolute Radio Frequency Channel Number}
\newacronym{ecaas}{ECaaS}{Edge-Cloud-as-a-Service}
\newacronym{ecn}{ECN}{Explicit Congestion Notification}
\newacronym{edf}{EDF}{Earliest Deadline First}
\newacronym{embb}{eMBB}{Enhanced Mobile Broadband}
\newacronym{empower}{EMPOWER}{EMpowering transatlantic PlatfOrms for advanced WirEless Research}
\newacronym{enb}{eNB}{evolved Node Base}
\newacronym{endc}{EN-DC}{E-UTRAN-\gls{nr} \gls{dc}}
\newacronym{epc}{EPC}{Evolved Packet Core}
\newacronym{eps}{EPS}{Evolved Packet System}
\newacronym{es}{ES}{Edge Server}
\newacronym{etsi}{ETSI}{European Telecommunications Standards Institute}
\newacronym[firstplural=Estimated Times of Arrival (ETAs)]{eta}{ETA}{Estimated Time of Arrival}
\newacronym{eutran}{E-UTRAN}{Evolved Universal Terrestrial Access Network}
\newacronym{faas}{FaaS}{Function-as-a-Service}
\newacronym{fapi}{FAPI}{Functional Application Platform Interface}
\newacronym{fdd}{FDD}{Frequency Division Duplexing}
\newacronym{fdm}{FDM}{Frequency Division Multiplexing}
\newacronym{fdma}{FDMA}{Frequency Division Multiple Access}
\newacronym{fed4fire}{FED4FIRE+}{Federation 4 Future Internet Research and Experimentation Plus}
\newacronym{fir}{FIR}{Finite Impulse Response}
\newacronym{fit}{FIT}{Future \acrlong{iot}}
\newacronym{fpga}{FPGA}{Field Programmable Gate Array}
\newacronym{fr2}{FR2}{Frequency Range 2}
\newacronym{fs}{FS}{Fast Switching}
\newacronym{fscc}{FSCC}{Flow Sharing Congestion Control}
\newacronym{ftp}{FTP}{File Transfer Protocol}
\newacronym{fw}{FW}{Flow Window}
\newacronym{ge}{GE}{Gaussian Elimination}
\newacronym{gnb}{gNB}{Next Generation Node Base}
\newacronym{llm}{LLM}{Large Language Model}
\newacronym{gop}{GOP}{Group of Pictures}
\newacronym{gpr}{GPR}{Gaussian Process Regressor}
\newacronym{gpu}{GPU}{Graphics Processing Unit}
\newacronym{gtp}{GTP}{GPRS Tunneling Protocol}
\newacronym{gtpc}{GTP-C}{GPRS Tunnelling Protocol Control Plane}
\newacronym{gtpu}{GTP-U}{GPRS Tunnelling Protocol User Plane}
\newacronym{gtpv2c}{GTPv2-C}{\gls{gtp} v2 - Control}
\newacronym{fwa}{FWA}{Fixed Wireless Access}
\newacronym{bic}{BIC}{Bayesian Information Criterion}
\newacronym{gw}{GW}{Gateway}
\newacronym{harq}{HARQ}{Hybrid Automatic Repeat reQuest}
\newacronym{hetnet}{HetNet}{Heterogeneous Network}
\newacronym{hh}{HH}{Hard Handover}
\newacronym{hol}{HOL}{Head-of-Line}
\newacronym{hqf}{HQF}{Highest-quality-first}
\newacronym{hss}{HSS}{Home Subscription Server}
\newacronym{http}{HTTP}{HyperText Transfer Protocol}
\newacronym{ia}{IA}{Initial Access}
\newacronym{iab}{IAB}{Integrated Access and Backhaul}
\newacronym{ic}{IC}{Incident Command}
\newacronym{ietf}{IETF}{Internet Engineering Task Force}
\newacronym{imsi}{IMSI}{International Mobile Subscriber Identity}
\newacronym{imt}{IMT}{International Mobile Telecommunication}
\newacronym{iot}{IoT}{Internet of Things}
\newacronym{ip}{IP}{Internet Protocol}
\newacronym{itu}{ITU}{International Telecommunication Union}
\newacronym{kpi}{KPI}{Key Performance Indicator}
\newacronym{kpm}{KPM}{Key Performance Measurement}
\newacronym{kvm}{KVM}{Kernel-based Virtual Machine}
\newacronym{los}{LoS}{Line of Sight}
\newacronym{lsm}{LSM}{Link-to-System Mapping}
\newacronym{lstm}{LSTM}{Long Short Term Memory}
\newacronym{lte}{LTE}{Long Term Evolution}
\newacronym{lxc}{LXC}{Linux Container}
\newacronym{m2m}{M2M}{Machine to Machine}
\newacronym{mac}{MAC}{Medium Access Control}
\newacronym{manet}{MANET}{Mobile Ad Hoc Network}
\newacronym{mano}{MANO}{Management and Orchestration}
\newacronym{mc}{MC}{Multi-Connectivity}
\newacronym{mcc}{MCC}{Mobile Cloud Computing}
\newacronym{mchem}{MCHEM}{Massive Channel Emulator}
\newacronym{mcs}{MCS}{Modulation and Coding Scheme}
\newacronym{mec2}{MEC}{Multi-access Edge Computing}
\newacronym{mec}{MEC}{Mobile Edge Computing}
\newacronym{mfc}{MFC}{Mobile Fog Computing}
\newacronym{mgen}{MGEN}{Multi-Generator}
\newacronym{mi}{MI}{Mutual Information}
\newacronym{mib}{MIB}{Master Information Block}
\newacronym{miesm}{MIESM}{Mutual Information Based Effective SINR}
\newacronym{mimo}{MIMO}{Multiple Input, Multiple Output}
\newacronym{ml}{ML}{Machine Learning}
\newacronym{mlr}{MLR}{Maximum-local-rate}
\newacronym[plural=\gls{mme}s,firstplural=Mobility Management Entities (MMEs)]{mme}{MME}{Mobility Management Entity}
\newacronym{mmtc}{mMTC}{Massive Machine-Type Communications}
\newacronym{mmwave}{mmWave}{millimeter wave}
\newacronym{mpdccp}{MP-DCCP}{Multipath Datagram Congestion Control Protocol}
\newacronym{mptcp}{MPTCP}{Multipath TCP}
\newacronym{mr}{MR}{Maximum Rate}
\newacronym{bdeu}{BDeU}{Bayesian Dirichlet equivalent uniform}
\newacronym{mrdc}{MR-DC}{Multi \gls{rat} \gls{dc}}
\newacronym{mse}{MSE}{Mean Square Error}
\newacronym{mss}{MSS}{Maximum Segment Size}
\newacronym{mt}{MT}{Mobile Termination}
\newacronym{mtd}{MTD}{Machine-Type Device}
\newacronym{mtu}{MTU}{Maximum Transmission Unit}
\newacronym{mumimo}{MU-MIMO}{Multi-user \gls{mimo}}
\newacronym{mvno}{MVNO}{Mobile Virtual Network Operator}
\newacronym{nalu}{NALU}{Network Abstraction Layer Unit}
\newacronym{nas}{NAS}{Non-Access Stratum}
\newacronym{nat}{NAT}{Network Address Translation}
\newacronym{nbiot}{NB-IoT}{Narrow Band IoT}
\newacronym{nfv}{NFV}{Network Function Virtualization}
\newacronym{nfvi}{NFVI}{Network Function Virtualization Infrastructure}
\newacronym{ni}{NI}{Network Interfaces}
\newacronym{nic}{NIC}{Network Interface Card}
\newacronym{now}{NOW}{Non Overlapping Window}
\newacronym{cpd}{CPD}{Conditional Probability Distribution}
\newacronym{nsm}{NSM}{Network Service Mesh}
\newacronym{nr}{NR}{New Radio}
\newacronym{nrf}{NRF}{Network Repository Function}
\newacronym{nsa}{NSA}{Non Stand Alone}
\newacronym{nse}{NSE}{Network Slicing Engine}
\newacronym{nssf}{NSSF}{Network Slice Selection Function}
\newacronym{o2i}{O2I}{Outdoor to Indoor}
\newacronym{oai}{OAI}{OpenAirInterface}
\newacronym{oaicn}{OAI-CN}{\gls{oai} \acrlong{cn}}
\newacronym{oairan}{OAI-RAN}{\acrlong{oai} \acrlong{ran}}
\newacronym{oam}{OAM}{Operations, Administration and Maintenance}
\newacronym{ofdm}{OFDM}{Orthogonal Frequency Division Multiplexing}
\newacronym{olia}{OLIA}{Opportunistic Linked Increase Algorithm}
\newacronym{omec}{OMEC}{Open Mobile Evolved Core}
\newacronym{onap}{ONAP}{Open Network Automation Platform}
\newacronym{onf}{ONF}{Open Networking Foundation}
\newacronym{onos}{ONOS}{Open Networking Operating System}
\newacronym{oom}{OOM}{\gls{onap} Operations Manager}
\newacronym{opnfv}{OPNFV}{Open Platform for \gls{nfv}}
\newacronym{oran}{O-RAN}{Open Radio Access Network}
\newacronym{orbit}{ORBIT}{Open-Access Research Testbed for Next-Generation Wireless Networks}
\newacronym{os}{OS}{Operating System}
\newacronym{oss}{OSS}{Operations Support System}
\newacronym{pa}{PA}{Position-aware}
\newacronym{pase}{PASE}{Prioritization, Arbitration, and Self-adjusting Endpoints}
\newacronym{pawr}{PAWR}{Platforms for Advanced Wireless Research}
\newacronym{pbch}{PBCH}{Physical Broadcast Channel}
\newacronym{pcef}{PCEF}{Policy and Charging Enforcement Function}
\newacronym{pcfich}{PCFICH}{Physical Control Format Indicator Channel}
\newacronym{pcrf}{PCRF}{Policy and Charging Rules Function}
\newacronym{pdcch}{PDCCH}{Physical Downlink Control Channel}
\newacronym{pdcp}{PDCP}{Packet Data Convergence Protocol}
\newacronym{pdsch}{PDSCH}{Physical Downlink Shared Channel}
\newacronym{pdu}{PDU}{Packet Data Unit}
\newacronym{pf}{PF}{Proportional Fair}
\newacronym{pgw}{PGW}{Packet Gateway}
\newacronym{phich}{PHICH}{Physical Hybrid ARQ Indicator Channel}
\newacronym{phy}{PHY}{Physical}
\newacronym{pmch}{PMCH}{Physical Multicast Channel}
\newacronym{pmi}{PMI}{Precoding Matrix Indicators}
\newacronym{powder}{POWDER}{Platform for Open Wireless Data-driven Experimental Research}
\newacronym{ppo}{PPO}{Proximal Policy Optimization}
\newacronym{ppp}{PPP}{Poisson Point Process}
\newacronym{prach}{PRACH}{Physical Random Access Channel}
\newacronym{prb}{PRB}{Physical Resource Block}
\newacronym{psnr}{PSNR}{Peak Signal to Noise Ratio}
\newacronym{pss}{PSS}{Primary Synchronization Signal}
\newacronym{pucch}{PUCCH}{Physical Uplink Control Channel}
\newacronym{pusch}{PUSCH}{Physical Uplink Shared Channel}
\newacronym{qam}{QAM}{Quadrature Amplitude Modulation}
\newacronym{qci}{QCI}{\gls{qos} Class Identifier}
\newacronym{qoe}{QoE}{Quality of Experience}
\newacronym{qos}{QoS}{Quality of Service}
\newacronym{quic}{QUIC}{Quick UDP Internet Connections}
\newacronym{ra}{RA}{Random Access}
\newacronym{rach}{RACH}{Random Access Channel}
\newacronym{ran}{RAN}{Radio Access Network}
\newacronym[firstplural=Radio Access Technologies (RATs)]{rat}{RAT}{Radio Access Technology}
\newacronym{rbg}{RBG}{Resource Block Group}
\newacronym{rcn}{RCN}{Research Coordination Network}
\newacronym{rc}{RC}{RAN Control}
\newacronym{rec}{REC}{Radio Edge Cloud}
\newacronym{red}{RED}{Random Early Detection}
\newacronym{renew}{RENEW}{Reconfigurable Eco-system for Next-generation End-to-end Wireless}
\newacronym{rf}{RF}{Radio Frequency}
\newacronym{rfc}{RFC}{Request for Comments}
\newacronym{rfr}{RFR}{Random Forest Regressor}
\newacronym{ric}{RIC}{RAN Intelligent Controller}
\newacronym{rlc}{RLC}{Radio Link Control}
\newacronym{rlf}{RLF}{Radio Link Failure}
\newacronym{rlnc}{RLNC}{Random Linear Network Coding}
\newacronym{rmr}{RMR}{RIC Message Router}
\newacronym{rmse}{RMSE}{Root Mean Squared Error}
\newacronym{rnis}{RNIS}{Radio Network Information Service}
\newacronym{rr}{RR}{Round Robin}
\newacronym{rrc}{RRC}{Radio Resource Control}
\newacronym{rrm}{RRM}{Radio Resource Management}
\newacronym{rru}{RRU}{Remote Radio Unit}
\newacronym{rs}{RS}{Remote Server}
\newacronym{rsrp}{RSRP}{Reference Signal Received Power}
\newacronym{rsrq}{RSRQ}{Reference Signal Received Quality}
\newacronym{rss}{RSS}{Received Signal Strength}
\newacronym{rssi}{RSSI}{Received Signal Strength Indicator}
\newacronym{rtt}{RTT}{Round Trip Time}
\newacronym{ru}{RU}{Radio Unit}
\newacronym{rw}{RW}{Receive Window}
\newacronym{rx}{RX}{Receiver}
\newacronym{s1ap}{S1AP}{S1 Application Protocol}
\newacronym{sa}{SA}{standalone}
\newacronym{sack}{SACK}{Selective Acknowledgment}
\newacronym{sap}{SAP}{Service Access Point}
\newacronym{sc2}{SC2}{Spectrum Collaboration Challenge}
\newacronym{scef}{SCEF}{Service Capability Exposure Function}
\newacronym{sch}{SCH}{Secondary Cell Handover}
\newacronym{scoot}{SCOOT}{Split Cycle Offset Optimization Technique}
\newacronym{sctp}{SCTP}{Stream Control Transmission Protocol}
\newacronym{sdap}{SDAP}{Service Data Adaptation Protocol}
\newacronym{sdk}{SDK}{Software Development Kit}
\newacronym{sdm}{SDM}{Space Division Multiplexing}
\newacronym{sdma}{SDMA}{Spatial Division Multiple Access}
\newacronym{sdn}{SDN}{Software-defined Networking}
\newacronym{sdr}{SDR}{Software-defined Radio}
\newacronym{seba}{SEBA}{SDN-Enabled Broadband Access}
\newacronym{sgsn}{SGSN}{Serving GPRS Support Node}
\newacronym{sgw}{SGW}{Service Gateway}
\newacronym{si}{SI}{Study Item}
\newacronym{sib}{SIB}{Secondary Information Block}
\newacronym{sinr}{SINR}{Signal to Interference plus Noise Ratio}
\newacronym{sip}{SIP}{Session Initiation Protocol}
\newacronym{siso}{SISO}{Single Input, Single Output}
\newacronym{sla}{SLA}{Service Level Agreement}
\newacronym{sm}{SM}{Service Model}
\newacronym{smo}{SMO}{Service Management and Orchestration}
\newacronym{smsgmsc}{SMS-GMSC}{\gls{sms}-Gateway}
\newacronym{snr}{SNR}{Signal-to-Noise-Ratio}
\newacronym{son}{SON}{Self-Organizing Network}
\newacronym{sptcp}{SPTCP}{Single Path TCP}
\newacronym{srb}{SRB}{Service Radio Bearer}
\newacronym{srn}{SRN}{Standard Radio Node}
\newacronym{srs}{SRS}{Sounding Reference Signal}
\newacronym{ss}{SS}{Synchronization Signal}
\newacronym{sss}{SSS}{Secondary Synchronization Signal}
\newacronym{st}{ST}{Spanning Tree}
\newacronym{svc}{SVC}{Scalable Video Coding}
\newacronym{tb}{TB}{Transport Block}
\newacronym{tcp}{TCP}{Transmission Control Protocol}
\newacronym{tdd}{TDD}{Time Division Duplexing}
\newacronym{tdm}{TDM}{Time Division Multiplexing}
\newacronym{tdma}{TDMA}{Time Division Multiple Access}
\newacronym{tfl}{TfL}{Transport for London}
\newacronym{tfrc}{TFRC}{TCP-Friendly Rate Control}
\newacronym{tft}{TFT}{Traffic Flow Template}
\newacronym{tgen}{TGEN}{Traffic Generator}
\newacronym{tip}{TIP}{Telecom Infra Project}
\newacronym{tm}{TM}{Transparent Mode}
\newacronym{to}{TO}{Telco Operator}
\newacronym{tr}{TR}{Technical Report}
\newacronym{trp}{TRP}{Transmitter Receiver Pair}
\newacronym{ts}{TS}{Technical Specification}
\newacronym{tti}{TTI}{Transmission Time Interval}
\newacronym{ttt}{TTT}{Time-to-Trigger}
\newacronym{tx}{TX}{Transmitter}
\newacronym{uas}{UAS}{Unmanned Aerial System}
\newacronym{uav}{UAV}{Unmanned Aerial Vehicle}
\newacronym{udm}{UDM}{Unified Data Management}
\newacronym{udp}{UDP}{User Datagram Protocol}
\newacronym{udr}{UDR}{Unified Data Repository}
\newacronym{ue}{UE}{User Equipment}
\newacronym{uhd}{UHD}{\gls{usrp} Hardware Driver}
\newacronym{ul}{UL}{Uplink}
\newacronym{um}{UM}{Unacknowledged Mode}
\newacronym{uml}{UML}{Unified Modeling Language}
\newacronym{upa}{UPA}{Uniform Planar Array}
\newacronym{upf}{UPF}{User Plane Function}
\newacronym{urllc}{URLLC}{Ultra Reliable and Low Latency Communications}
\newacronym{usa}{U.S.}{United States}
\newacronym{usim}{USIM}{Universal Subscriber Identity Module}
\newacronym{usrp}{USRP}{Universal Software Radio Peripheral}
\newacronym{utc}{UTC}{Urban Traffic Control}
\newacronym{vim}{VIM}{Virtualization Infrastructure Manager}
\newacronym{vm}{VM}{Virtual Machine}
\newacronym{vnf}{VNF}{Virtual Network Function}
\newacronym{volte}{VoLTE}{Voice over \gls{lte}}
\newacronym{voltha}{VOLTHA}{Virtual OLT HArdware Abstraction}
\newacronym{vr}{VR}{Virtual Reality}
\newacronym{vran}{vRAN}{Virtualized \gls{ran}}
\newacronym{vss}{VSS}{Video Streaming Server}
\newacronym{wbf}{WBF}{Wired Bias Function}
\newacronym{wf}{WF}{Waterfilling}
\newacronym{wg}{WG}{Working Group}
\newacronym{wlan}{WLAN}{Wireless Local Area Network}
\newacronym{osm}{OSM}{Open Source \gls{nfv} Management and Orchestration}
\newacronym{pnf}{PNF}{Physical Network Function}
\newacronym{drl}{DRL}{Deep Reinforcement Learning}
\newacronym{mtc}{MTC}{Machine-type Communications}
\newacronym{osc}{OSC}{O-RAN Software Community}
\newacronym{mns}{MnS}{Management Services}
\newacronym{ves}{VES}{\gls{vnf} Event Stream}
\newacronym{ei}{EI}{Enrichment Information}
\newacronym{fh}{FH}{Fronthaul}
\newacronym{fft}{FFT}{Fast Fourier Transform}
\newacronym{laa}{LAA}{Licensed-Assisted Access}
\newacronym{plfs}{PLFS}{Physical Layer Frequency Signals}
\newacronym{ptp}{PTP}{Precision Time Protocol}
\newacronym{lidar}{LiDAR}{Light Detection And Ranging}
\newacronym{dem}{DEM}{Digital Elevation Model}
\newacronym{dtm}{DEM}{Digital Terrain Model}
\newacronym{dsm}{DEM}{Digital Surface Models}
\newacronym{ota}{OTA}{Over-The-Air}
\newacronym{ns}{NS}{Network Slicing}
\newacronym{ne}{NE}{Nash Equilibrium}
\newacronym{hf}{HF}{High Frequency}
\newacronym{noma}{NOMA}{Non-Orthogonal Multiple Access}
\newacronym{sre}{SRE}{Smart Radio Environment}
\newacronym{ris}{RIS}{Reconfigurable Intelligent Surface}
\newacronym{inp}{InP}{Infrastructure Provider}
\newacronym{smf}{SMF}{Slicing Magangement Framework}
\newacronym{nsn}{NSN}{Network Slicing Negotiation}
\newacronym{sms}{SMS}{Slicing MAC Scheduler}
\newacronym{brd}{BRD}{Best Response Dynamics}
\newacronym{dssbr}{DSSBR}{Double Step Smoothed Best Response}
\newacronym{poa}{PoA}{Price of Anarchy}
\newacronym{pos}{PoS}{Price of Stability}
\newacronym{milp}{MILP}{Mixed Integer-Linear Program}
\newacronym{pod}{PoD}{Price of DSSBR}
\newacronym{roc}{ROC}{Radio Overload Control}
\newacronym{ciot}{cIoT}{critical Internet of Things}
\newacronym{embbpr}{eMBB Pr.}{enhanced Mobile BroadBand Premium}
\newacronym{embbbs}{eMBB Bs.}{enhanced Mobile BroadBand Basic}
\newacronym{en}{EN}{Edge Node}
\newacronym{ec}{EC}{Edge Computing}
\newacronym{sp}{SP}{Service Provider}
\newacronym{me}{ME}{Market Equilibrium}
\newacronym{so}{SO}{Social Optimum}
\newacronym{wso}{WSO}{Weighted Social Optimum}
\newacronym{wsn}{WSN}{Wireless Sensor Network}
\newacronym{ps}{PS}{Proportional Sharing}
\newacronym{eg}{EG}{Eisenberg-Gale program}
\newacronym{pe}{PE}{Pareto Efficiency}
\newacronym{nsw}{NSW}{Nash Social Welfare}
\newacronym{ef}{EF}{Envy-Freeness}
\newacronym{sub6}{sub6GHz}{Below 6GHz}
\newacronym{ncr}{NCR}{Network-Controlled Repeater}
\newacronym{nlos}{NLoS}{Non-LoS}
\newacronym{src}{SRC}{Smart Radio Connection}
\newacronym{srd}{SRD}{Smart Radio Device}
\newacronym{cs}{CS}{Candidate Site}
\newacronym{tp}{TP}{Test Point}
\newacronym{fov}{FoV}{Field of View}
\newacronym{nrric}{Near-RT RIC}{Near Real-time RAN Intelligent Controller}
\newacronym{e2ap}{E2AP}{E2 Application Protocol}
\newacronym{e2sm}{E2SM}{E2 Service Model}
\newacronym{nrtric}{Non-RT RIC}{Non-Real-Time Ran Intelligent Controller}
\newacronym{itti}{ITTI}{Inter-task Interface}
\newacronym{bap}{BAP}{Backhaul Adaptation Protocol}
\newacronym{iabest}{IABEST}{Integrated Access and Backhaul Experimental large-Scale Tetbed}
\newacronym{teid}{TEID}{Tunnel Endpoint Identifier}
\newacronym{dlsch}{DL-SCH}{Downlink Shared Channel }
\newacronym{ulsch}{UL-SCH}{Uplink Shared Channel }
\newacronym{opex}{OpEx}{Operational Expenditure}
\newacronym{capex}{CapEx}{Capital Expenditure}
\newacronym{mno}{MNO}{Mobile Network Operator}
\newacronym{fr}{FR}{Frequency Range}
\newacronym{bn}{BN}{Bayesian Network}

\def\BibTeX{{\rm B\kern-.05em{\sc i\kern-.025em b}\kern-.08em
    T\kern-.1667em\lower.7ex\hbox{E}\kern-.125emX}}

\def\framework{RANalyzer\xspace}

\begin{document}

\title{\framework: Automated Continuous RAN Software Evaluation and Regression Analysis
\thanks{This paper is based on material supported in part by the U.S. National Telecommunications and Information Administration (NTIA)'s Public Wireless Supply Chain Innovation Fund (PWSCIF) under Award No. 25-60-IF054 and NSF TI-2449452.}
\thanks{\IEEEauthorrefmark{1}The authors contributed equally.}
}

\author{\IEEEauthorblockN{Ravis Shirkhani\IEEEauthorrefmark{1}, 
Reshma Prasad\IEEEauthorrefmark{1}, 
Leonardo Bonati, 
Tommaso Melodia,
Michele Polese}
\IEEEauthorblockN{
Institute for Intelligent Networked Systems, Northeastern University, Boston, MA, USA\\
Email: \{shirkhani.r, re.prasad, l.bonati, melodia, m.polese\}@northeastern.edu
}}

\maketitle

\ifnumequal{\thepage}{1}{%
    \begin{tikzpicture}[remember picture,overlay]
        \node[draw,
        minimum width=0.6\paperwidth,
        text width=0.6\paperwidth,
        align=center,
        font=\scriptsize,
        anchor=north
        ]
        at ($(current page.north)+(0,-15pt)$)
        {%
        This article has been accepted for publication in IEEE International Conference on Network Softwarization (NetSoft 2026)
        };
    \end{tikzpicture}%
}{}

\begin{abstract}
Software-driven O-RAN architectures enable rapid innovation through frequent, independent updates to virtualized components. However, attributing performance variations to specific software changes is challenging due to the stochastic nature of wireless systems, where channel conditions, interference, and hardware variability confound analysis. Traditional threshold-based monitoring and manual troubleshooting do not scale with modern software evolution.

This paper presents \framework, an automated test analysis framework that quantifies the performance impact of software updates beyond what can be explained by wireless channel conditions. 
\framework combines LLM-assisted semantic extraction with residuals analysis. The first categorizes code changes by affected protocol layers and functional components, while the second provides insights on the effect of load, channel, or code changes on the test performance.
We contribute an extensive dataset collected over more than two years of continuous over-the-air testing on an experimental O-RAN testbed, comprising over 8,600 automated tests across 69 releases of the \gls{oai} stack. By modeling expected performance and interpreting deviations as software-induced effects, we identify degraded instances attributable to code changes and correlate them with specific change categories. The framework can be integrated into CI/CD/CT pipelines for automated, continuous evaluation of software updates at scale.

\end{abstract}


\glsresetall
\glsunset{nr}

\section{Introduction}
\label{introduction}

\gls{ran} architectures are increasingly software-driven, programmable, and disaggregated, driven by 3GPP \gls{nr} and O-RAN ALLIANCE specifications~\cite{abdalla2022toward,polese2023understanding,agarwal2025open}. A growing portion of the network is implemented in software, including virtualized \glspl{cu} and \glspl{du}, \glspl{ric}, and their hosted control and optimization applications, such as rApps, xApps, and dApps. This softwarization enables flexible deployment and creates opportunities for rapid innovation and fine-grained network control. At the same time, tight coupling to a complex software ecosystem introduces new challenges for network design, implementation, operation, and performance. In particular, attributing observed performance variations to individual software updates becomes difficult as updates occur frequently and span independently developed components.

In response to this software complexity, automated frameworks for \gls{ci}, \gls{cd}, and \gls{ct} have been introduced to integrate updates, deploy, test, and validate cellular network software at scale~\cite{luis_herrera_tutorial_2026, motamary_deep_2023}. 
Such frameworks 
also enable continuous \gls{ota} testing of 
\gls{ran} components under realistic channel conditions using experimental testbeds, and detailed collection of test results over time, e.g., from network logs and traffic-generation tools~\cite{bonati20245g}. 
The research and prototyping on these frameworks, however, has primarily focused on data collection and performance validation, while providing limited support for systematic performance analysis at scale and for the interpretation of performance variations across tests.

Performance analysis is commonly performed using threshold-based monitoring of \glspl{kpi}, comparisons against historical \glspl{kpi}, and post-deployment manual troubleshooting when regressions are suspected~\cite{zelalem_jembre_mobile_2022}. While these approaches are effective for detecting service-level objective failures, they rely on coarse-grained metrics and assume infrequent and limited software updates. However, in O-RAN deployments, multiple software components are updated independently and at a much faster pace (e.g., as fast as on a weekly basis). As a result, manual inspection of test results and simple statistical comparisons across software versions do not scale and provide limited insight into how individual code changes affect network performance. Furthermore, unit tests alone cannot capture the complex cross-component interactions that arise in this disaggregated architecture. 
\vspace{-1em}
\begin{figure}[h]
  \centering
  \includegraphics[width=0.95\columnwidth]{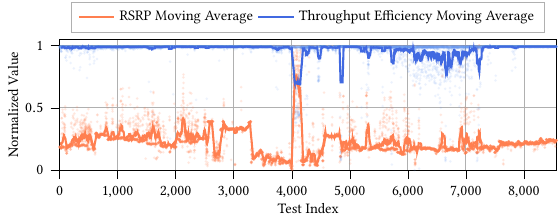}
          \vspace{-.3cm}
   \caption{Moving average of downlink RSRP and throughput efficiency across all tests considered. High signal strength does not consistently
correspond to high throughput efficiency.}
    \label{fig:rsrp-throughput}
        \vspace{-.5cm}
\end{figure}

\begin{figure*}[t]
\setlength\abovecaptionskip{1pt}
    \centering
    \includegraphics[width=0.9\textwidth,keepaspectratio]{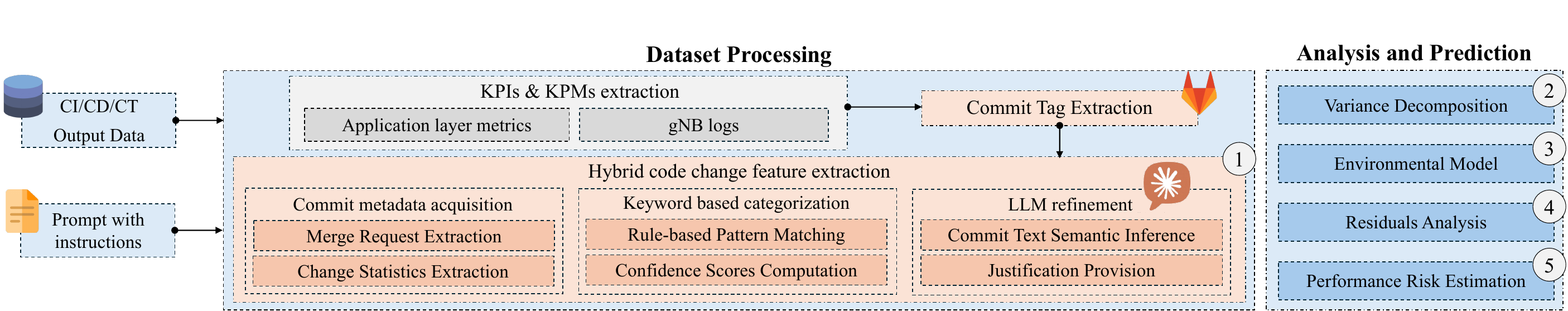}
        \setlength{\belowcaptionskip}{-.3cm}
    \caption{High-level overview of \framework.}
    \label{fig:system-model}
\end{figure*}

This challenge is further exacerbated by the stochastic nature of wireless systems, where performance is inherently affected by channel conditions, interference, and hardware variability. As a result, tests executed with identical software configurations may yield different performance outcomes and can mask the true impact of software changes. 

To illustrate this, Figure~\ref{fig:rsrp-throughput} reports the moving average of throughput efficiency, defined as the ratio of actual to target throughput, and downlink \gls{rsrp} of daily tests for the open-source \gls{oai} protocol stack over more than 24 months. While higher \gls{rsrp} generally indicates favorable channel conditions, the figure reveals groups of tests where throughput efficiency remains low despite consistently strong signal strength. This mismatch suggests contribution of other factors on performance variations beyond channel conditions.
The growing mismatch between software evolution pace and analysis capabilities, combined with the challenge of distinguishing software-induced performance changes from environmental factors, necessitates automated and systematic performance analysis methodologies that can account for multiple confounding factors.

\textbf{Contributions.} To address this gap, in this paper we design, prototype, and evaluate \framework, an automated test analysis solution for software-driven cellular networks. 
\framework quantifies the performance impact of software updates in O-RAN systems, beyond what can be explained by wireless channel conditions. 
Figure~\ref{fig:system-model} shows a high-level view of \framework. It combines (i) a \gls{llm}-assisted method to automatically categorize software updates by extracting semantic information from commit messages (ii) stochastic analysis that accounts for confounding elements and test conditions.

To design \framework, we leverage an extensive continuous \gls{ota} testing dataset, collected over more than two years on an experimental O-RAN testbed~\cite{bonati20245g}. It includes \gls{e2e} performance metrics and detailed \gls{gnb} logs across 69 software releases of \gls{oai}. 
We show the effectiveness of reasoning capabilities of \glspl{llm} in understanding semantics over unstructured text (e.g., commit messages). 
We then model the expected network performance under given environmental and load conditions, and interpret deviations from this expected behavior as indicators of software-induced performance improvements or degradations. By combining this residual-based analysis with semantic features extracted from software commits, we provide systematic insights into how different classes of code changes affect mobile network performance. Once trained, the proposed framework can be integrated into any \gls{ci}/\gls{cd}/\gls{ct} pipeline to enable automated testing and evaluation of software updates at scale, allowing new software releases to be assessed continuously as part of the workflow.

The contributions of the paper are as follows:
\begin{itemize}
    \item We collect and analyze a dataset obtained through continuous \gls{ota} testing over more than two years, comprising more than 8600 automated tests with \gls{e2e} performance metrics and detailed \gls{gnb} logs across a wide range of \gls{oai} protocol stack updates and traffic loads.\footnote{The collected dataset is available at https://github.com/wineslab/RANalyzer-Dataset.}

    \item We introduce a hybrid code-change characterization approach that combines keyword-based pattern matching with \gls{llm}-assisted refinement to automatically extract and classify semantic features from software commits, including affected protocol layers and functional components.
    
    \item We propose an automated methodology to isolate code-induced performance effects from other sources of variability by explicitly accounting for traffic load and wireless channel conditions.

    \item We demonstrate \framework on real-world 5G \gls{ran} tests, identifying  degraded instances attributed to code-related issues. We validate the practical utility of our approach through case study analysis correlating degradations with code change categories. 
\end{itemize}

The remainder of this paper is organized as follows. In Section~\ref{realted-works}, we review related literature works. In Section~\ref{system-description}, we provide a high-level description of the workflows on the 5G infrastructure. In Section~\ref{data-processing}, we provide details on the collected dataset. In Section~\ref{sec:framework-workflow}, we propose \framework and analyze code-induced performance changes. Finally, we discuss our results in Section~\ref{sec:evaluation}, and draw our conclusions in Section~\ref{sec:conclusion}.

\section{Related Work}
\label{realted-works}

Detection of performance degradation  in software changes has been extensively studied, but existing approaches face significant challenges when applied to wireless systems where performance depends on both code quality and environmental factors.
Previous statistical methods, designed for stable environments like 
data centers, include controlled comparison approaches. The authors of~\cite{zhang2015rapid} employ Difference-in-Differences (DiD) analysis by comparing performance in servers with old and new software versions simultaneously.  Authors of~\cite{yu2024changerca} extend this with two-layer DiD to account for seasonality, comparing post-change instances against both pre-change instances and historical baselines.  DiD methods~\cite{zhang2015rapid, yu2024changerca} require  identical environmental conditions across versions, an assumption that fails in wireless cellular testing~\cite{mahimkar2013robust}. 
Indeed, channel conditions may vary,
causing performance differences that would be incorrectly attributed to code changes under DiD analysis. \cite{li2020gandalf} uses ensemble voting and temporal-spatial correlation to analyze failures over specific components among hundreds of concurrent rollouts in Azure infrastructure, as well as historical human-labeled deployment data to decide whether to proceed with the rollout or not. 
Beyond statistical methods, machine learning approaches have been explored. The authors of~\cite{zhao2021identifying} treat  defect identification as an anomaly detection task. This approach  monitors performance data from multiple source and utilizes a multivariate LSTM  to capture defective changes.
\cite{wang2022identifying} uses distance-based comparison on \gls{kpi} time-series, comparing post-change measurements against historical and pre-change baselines via Siamese LSTM. However, this assumes environmental stability, which is valid for data centers but fails in wireless networks where channel conditions can vary over time.

Performance degradation analysis in cellular networks has been explored in related contexts.
\cite{mahimkar2013robust} employs spatial regression to assess configuration changes by comparing study group against control group (changed and unchanged deployments). This approach learns pre-change relationships via linear regression, then forecasts expected post-change performance from control group behavior. 
This requires similar unchanged deployments as predictors, whereas we use environmental features directly, eliminating control group requirements. The authors of~\cite{patel2024cipat,patel2024predicting} propose the use of \gls{codec}, a variance-based statistical method, to identify which configuration parameters affect performance. They build a two-stage model 
that predicts performance impact, identifying parameters using \gls{codec} scores first, then building predictive models for continuous parameter optimization. Their approach requires continuous, directional variables for predicting change direction, but code changes are categorical. In our framework, we use a similar variance decomposition approach to quantify the contribution of various factors to performance.

Network change analysis has been addressed through pre/post statistical comparison methods including Kruskal-Wallis testing~\cite{mahimkar2022aurora}
and rank-based CUSUM for upgrade detection~\cite{mahimkar2010detecting}. These approaches detect performance shifts but cannot distinguish code-induced changes from environmental variations.

\section{System Description}
\label{system-description}

This section provides a high-level description of the system and workflows used for this work. We introduce the automated private 5G infrastructure and the \gls{ci}/\gls{cd}/\gls{ct} workflows used to continuously deploy and test \gls{ran} software. We then describe the system design of \framework.

\subsection{Automated 5G Infrastructure}
\label{sec:5g-infrastructure}
We leverage a heterogeneous private 5G network testbed designed to support automated, repeatable, and large-scale experimentation. The testbed is deployed on an OpenShift cluster and integrates open-source and commercial components (e.g., core networks, multiple \glspl{ru} and \glspl{sdr}) for a complete \gls{e2e} 5G network.

Network functions are deployed as containerized workloads on general-purpose compute nodes (e.g., the \gls{ran} uses Microway servers with AMD EPYC 7262). OpenShift provides native support for application lifecycle management, container orchestration, and automation primitives that are essential for continuous experimentation. It integrates \gls{ci}/\gls{cd} capabilities through cloud-native tools such as Tekton and GitOps-based configuration management via ArgoCD. 

The \gls{ran} workloads deployed on the cluster interface with a grid of \glspl{sdr} (USRP X410)
to enable \gls{ota} tests in a scattering-rich indoor space. The \glspl{ue} used in the testbed are commercial Sierra Wireless 5G modems connected to a small host compute node, enabling automated control of attachment procedures and traffic generation. Open5Gs is used as core network. Together, the containerized network functions, radio grid, and commercial \glspl{ue} form the physical and virtual foundation on which the automated \gls{ci}/\gls{cd}/\gls{ct} workflows described in the following subsection are implemented.

\subsection{\gls{ci}/\gls{cd}/\gls{ct} Workflow on the Testbed}
\label{subsec:testbed-workflow}

\begin{figure}[h]
  \centering
  \includegraphics[width=0.95\columnwidth]{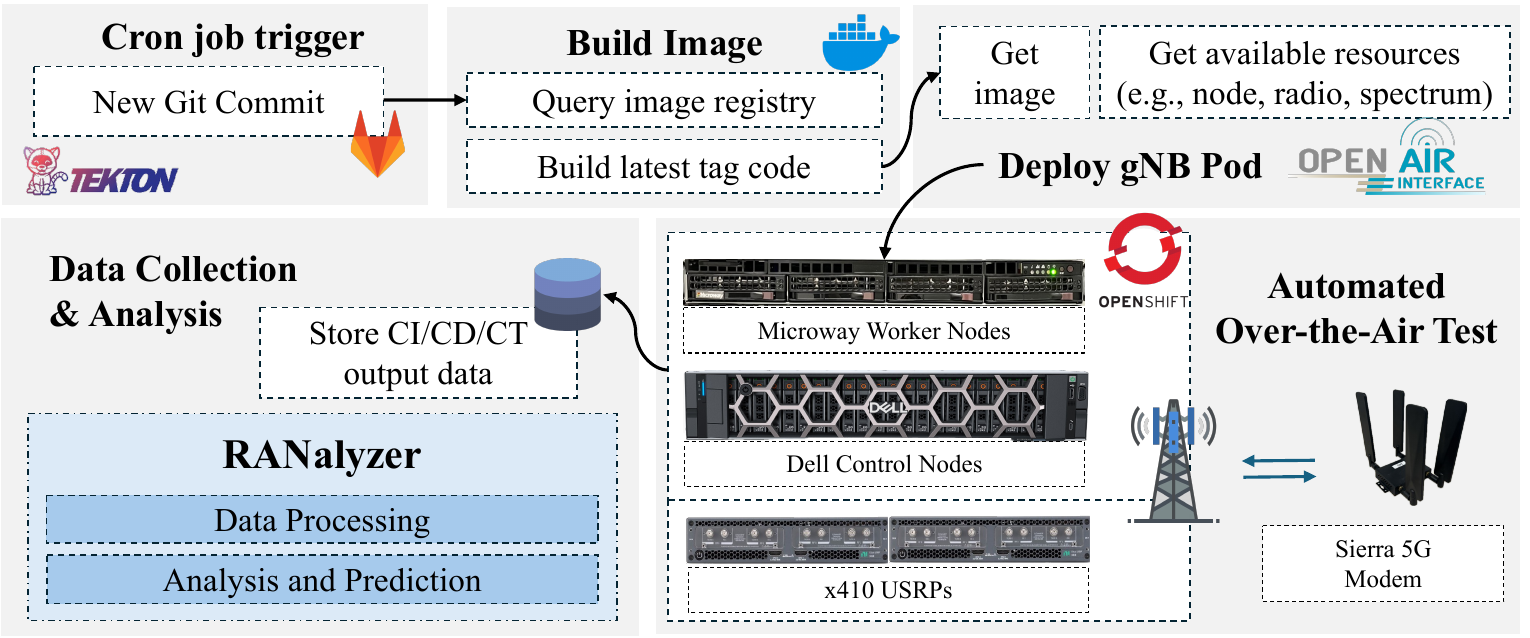}
   \caption{High-level overview of \gls{ci}/\gls{cd} and \framework workflow on the 5G infrastructure.}
    \label{fig:cicd-workflow}
        \vspace{-.3cm}
\end{figure}

Through the \gls{ci}/\gls{cd}/\gls{ct} capabilities available on the Openshift cluster, we collect our dataset with the 5G-CT framework described in~\cite{bonati20245g}. In this framework, automated pipelines build, deploy, and test an \gls{e2e} network through series of tasks. Figure~\ref{fig:cicd-workflow} shows the high-level components of 5G-CT. The sequence of tasks includes periodic triggers, retrieving the latest \gls{oai} software tag from GitLab, checking previously built images in a registry, building a new image when a new tag is detected, deploying the workload on the mentioned Microway, running continuous tests \gls{ota} with constant bandwidth of $60$MHz over carrier frequency $3.629$\:GHz in band n78, and finally storing the collected artifacts in a database. \cite{maxenti_autoran_2025} provides the means and configurations to have this deployment and testing on the Openshift cluster. This design allows the \gls{gnb} software to be continuously updated while keeping test execution repeatable across software versions. Using this methodology, we collected a dataset that spans 69 releases of \gls{oai} from July 2023 to December 2025, with more than 8600 automated tests.
\vspace{-1em}

Each \gls{e2e} test runs downlink traffic using iPerf at multiple target data rates (e.g., 10, 20, and 30~Mbps, as well as higher rates such as 80~Mbps). At the end of each test, \gls{gnb} logs, iPerf results, and \gls{ue} metadata (e.g., serial number, hostname) are stored in a database for further analysis. 

\subsection{Introduction to \framework}
\label{sec:intro-to-framework}

\framework is an analysis layer designed to operate alongside the \gls{ci}/\gls{cd}/\gls{ct} workflow. As illustrated in Figure~\ref{fig:cicd-workflow}, \framework interfaces directly with the automated pipeline and consumes the artifacts produced and stored by the tests. The framework augments the existing testing infrastructure without modifying deployment or execution procedures, but by providing key insights into the stack performance in the tests.

At a high level, \framework aggregates results across software versions and test conditions, processes raw artifacts (test results, commit messages, stack logs) into structured inputs, and supports automated analysis of performance behavior under continuous updates by series of steps spanning through mathematical variance analysis and training ML models. Once trained, \framework can be integrated into \gls{ci}/\gls{cd}/\gls{ct} pipelines to enable automated analysis of test results, e.g., anytime there is a software update. Next, we describe the dataset built from these \gls{ota} tests, as its structure is essential for understanding the subsequent data processing and analysis steps.

\section{\framework Dataset}
\label{data-processing}

This section describes the \framework dataset, comprising two years of results of continuous and automated \gls{ota} tests. We discuss how we transform the raw output of the \gls{ota} tests into an analysis-ready dataset and analyze exemplary metrics.

\subsection{\gls{ci}/\gls{cd}/\gls{ct} Output Data Organization}
\label{dataset-organization}

Data is organized into folders based on the date and time the automated pipeline ran each test (triggered via cron jobs on OpenShift). At the top level, each day corresponds to a single directory with name in \texttt{yyyymmdd} format (e.g., \texttt{20250913/}). Within each day, multiple tests are executed, each one stored in a separate subdirectory based on its start time in the \texttt{hhmmss} format (e.g., \texttt{040124/}). Each test directory contains: (i) iPerf3 results saved as \texttt{csv} files with \gls{e2e} performance metrics (throughput, packet loss, and jitter); and (ii) \gls{oai} \gls{gnb} runtime logs from which \glspl{kpm} and protocol-level events can be extracted.

\subsection{\gls{kpi}/\gls{kpm}  and Dataset Consolidation}
\label{sec:kpi-kpm-features}
To facilitate the parsing of the dataset for further analysis, we convert the heterogeneous result files from the \gls{ota} tests (e.g., log files and iPerf results) into a unified set of numerical features aligned at the test level.

\textbf{\gls{e2e} \glspl{kpi} from iPerf.} We parse the iPerf output to extract application-level metrics including throughput, packet loss, jitter, and total bytes and packets. Because tests are executed with different target rates (ranging from $10$\:Mbps to $80$\:Mbps), we also compute normalized metrics that enable comparison across loads, such as throughput efficiency:
\begin{equation}
\eta_{\text{test}} = T_{\text{test}}/T_{\text{target}}
\end{equation}
where $T_{\text{test}}$ denotes the average measured throughput and $T_{\text{target}}$ the requested data rate.

\textbf{Low-level \gls{kpm} and event metrics from gNB logs.} In addition to \gls{e2e} \glspl{kpi}, we extract radio and protocol indicators from the \gls{gnb} logs. These include numerical \glspl{kpm} reported across multiple layers of the \gls{ran} stack, such as physical-layer measurements (e.g., RSRP, SINR), link-layer reliability indicators (e.g., BLER and HARQ retransmission statistics), and scheduling- and control-related metrics from higher layers (e.g., CQI reports). In parallel, we extract event-based features corresponding to protocol and system events across the PHY, MAC, and RRC layers, including connection setup and release markers, scheduling anomalies, warnings, and error patterns (e.g., number of active \gls{pdu} sessions, failed msg2 in the \gls{ra} procedure). Together, these numerical and event-based log features provide a comprehensive characterization of the radio environment and protocol behavior during each test.

\textbf{Dataset assembly and identifiers.} For the unified dataset construction, the extracted \glspl{kpi}/\glspl{kpm} from each test are aggregated into a single vector, indexed by a unique test identifier (timestamp-based folder name) and linked to the software revision identifier (commit hash). Over 80 features are extracted in this step. These features are combined with code-related features, extracted from the corresponding commit hash, as described in Section~\ref{sec:hybrid}.

While this Section focuses on offline post-processing of all the collected \gls{ci}/\gls{cd}/\gls{ct} test outputs, the same aggregation and feature-extraction can be executed automatically at the end of an individual test run. Specifically, a dedicated Python data-processing component runs inside an OpenShift pod upon test completion to collect the generated logs and measurement artifacts at the last step before \framework (Figure~\ref{fig:cicd-workflow}). The feature-extraction pipeline and the rest of the \framework workflow can be executed as subsequent tasks within the same pod, enabling a complete \gls{e2e} execution and analysis workflow.

\vspace{-1em}
\begin{figure*}[t]
  \centering
  \begin{subfigure}[t]{0.32\textwidth}
    \centering
    \includegraphics[width=\linewidth]{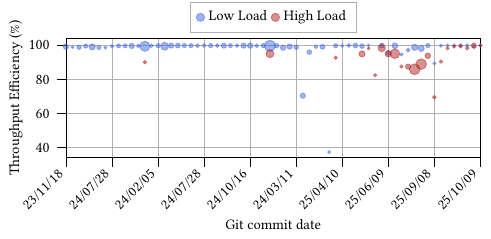}
    \caption{Throughput efficiency under low and high load}
    \label{fig:throughput-load}
  \end{subfigure}
  \hfill
  \begin{subfigure}[t]{0.32\textwidth}
    \centering
    \includegraphics[width=\linewidth]{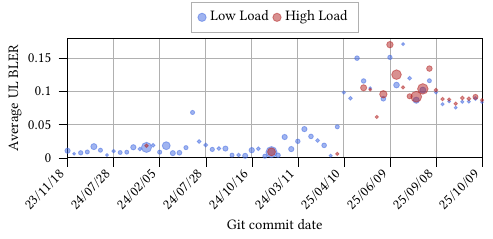}
    \caption{UL BLER under low and high load}
    \label{fig:ulbler-load}
  \end{subfigure}
  \hfill
  \begin{subfigure}[t]{0.32\textwidth}
    \centering
    \includegraphics[width=\linewidth]{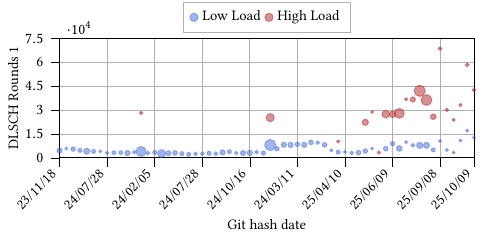}
    \caption{DL-SCH transport block retransmissions}
    \label{fig:harq1-load}
  \end{subfigure}
  \caption{Metrics across Git commits for low- and high-load test cases.}
  \label{fig:performance-load-comparison}
\end{figure*}

\subsection{Dataset Analysis}
\label{sec:dataset-analysis}

We perform an analysis on different \glspl{kpi} and \glspl{kpm} of our dataset to show how different factors influence system performance. In particular, we show that performance is strongly affected by factors such as channel conditions and traffic load.
To understand the impact of traffic load, we separately analyze tests with high load (${\ge}50$\:Mbps) and low load (${<}50$\:Mbps).\footnote{High or low load depend on the capacity of the infrastructure on which the tests are executed and test bandwidth.} 
Figure~\ref{fig:throughput-load} shows the throughput efficiency across different Git tags (i.e., a new release) for low-load (blue curve) and high-load (red curve) tests. The marker size is proportional to the number of tests executed for each Git tag. While performance fluctuates across Git tags, the overall trend shows that higher-load tests tend to achieve lower throughput efficiency. This observation further highlights the importance of accounting for traffic load in the performance analysis. Figure~\ref{fig:ulbler-load} shows the average uplink BLER, extracted from the gNB logs. For this metric, we observe an overall increase toward the more recent commits compared to earlier ones. The values are similar for both load levels, which suggests that the observed changes are likely driven by environmental factors, such as channel conditions, which affect all tests in a similar way. Another interesting feature is the number of first round \gls{harq} retransmissions, shown in Figure~\ref{fig:harq1-load}.  We observe that the values increase over time but remain consistently higher under high load, as expected from aggressive scheduling and higher modulation schemes under traffic pressure.

\section{\framework Automated Workflow}
\label{sec:framework-workflow}

Analysis in the previous sections showed that performance variations in wireless systems arise from multiple interacting factors, including environmental conditions (channel quality and interference), workload characteristics (traffic load), and code modifications (protocol-layer changes). We develop a multi-stage framework to identify and attribute performance changes to specific causes. As shown at a high-level in Figure~\ref{fig:system-model}, the framework consists of five stages: (i) automated hybrid code change categorization, (ii) variance decomposition, (iii) environment modeling, (iv) residual-based performance analysis, and (v) performance risk estimation. In this section, we discuss each of these stages in detail.

\subsection{Automated Hybrid Code Change Categorization}
\label{sec:hybrid}

 In the first step of \framework, we automatically extract a structured feature vector from each \gls{oai} commit to relate performance variations to software evolution. For a given commit $c$, we retrieve its message from GitLab, along with basic code-churn (e.g., files changed and lines added/deleted), and aggregate this information into a single text record $m_c$. We then infer protocol layers and functional components that are impacted by the change and assign a high-level change type (e.g., bug fix, optimization, feature, refactoring).

We propose a hybrid feature extraction pipeline that combines fast keyword-based categorization with selective \gls{llm}-based semantic refinement, enabling scalable and accurate processing of large commit histories with ambiguous terminology. Let $\mathcal{L}$ denote the set of protocol layers (PHY, MAC, RLC, PDCP, RRC, NAS, NGAP, F1AP, E1AP) and $\mathcal{C}$ the set of functional components (memory, threading, radio, scheduler, timer, queue). We define $\Gamma = \mathcal{L} \cup \mathcal{C}$, with $k = |\Gamma|$. For a commit $c$, the affected categories $\Gamma_c \subseteq \Gamma$ are encoded as a binary feature vector $\mathbf{f}_c \in \{0,1\}^k$, where $\mathbf{f}_c[i]=1$ if $c$ affects category $i$, and $0$ otherwise (e.g., for $\Gamma_c=\{\text{PHY},\text{MAC}\}$, $\mathbf{f}_c=[1,1,0,\ldots,0]$). This vector is populated via the following automated steps:

\subsubsection{Keyword-Based Categorization}
\label{sec:keyword}

The first stage of the hybrid pipeline performs fast, rule-based categorization using domain-specific keyword sets organized by protocol layer and functional component. The keyword sets were constructed through expert-guided curation based on \gls{ran} protocol specifications and the terminology of the \gls{oai} codebase. The input to this stage is the aggregated commit text $m_c$, constructed from the commit title, description, and merge-request references extracted from the GitLab commit page.

For each commit, we apply case-insensitive substring matching against predefined keyword sets associated with each protocol layer and component. Each set contains characteristic terms commonly used in that part of the \gls{ran} stack (e.g., \verb|L1| and \verb|NR| for PHY, \verb|MAC| and \verb|LogicalChannelConfig| for MAC, and \verb|RRC| for the RRC layer). Keywords are labeled as strong, medium, or weak, reflecting their specificity (e.g., the term \verb|NR_MAC| is strong, while \verb|scheduler| is weak as it may appear across multiple layers), and contribute weighted evidence to the corresponding category. The weighted keyword score for a given category $j$ is computed as
\begin{equation}
\sum_{i \in \mathcal{K}_j} w_i \cdot \mathbf{1}_{\text{match}}(i),
\end{equation}
where $\mathcal{K}_j$ denotes the keyword set for category $j \in \Gamma$, $w_i \in \{2, 1, 0.5\}$ corresponding to strong, medium, and weak keywords, and $\mathbf{1}_{\text{match}}(i)$ indicates whether keyword $i$ appears in $m_c$. A category is marked as affected if its score exceeds a category-specific threshold $\theta_j$.
This procedure yields a binary feature vector indicating the set of affected layers and components, as well as aggregate statistics such as the total number of detected layers $L$ and components $C$.

In addition, the keyword-based stage assigns an extraction confidence label using a deterministic, rule-based logic. We denote this dependency abstractly as
\begin{equation}
\text{Confidence} \triangleq \mathcal{R}(L, C, K, S),
\end{equation}
where $\mathcal{R}(\cdot)$ represents a threshold-based decision function. The confidence depends on four quantities: the number of detected protocol layers $L$ and functional components $C$, the total amount of keyword evidence $K$, and the number of matched strong keywords $S$. Commits with large $K$ and $S$ concentrated within a small $L$ and $C$ are labeled as high confidence, while commits exhibiting weak or diffuse evidence across many categories are labeled as medium or low confidence. This confidence label is rule-based rather than probabilistic and is used to gate subsequent semantic \gls{llm}-refinement.

Figure~\ref{fig:keyword_categorization} illustrates representative commit messages and their matched keyword patterns, highlighting how layer-specific terminology enables rapid identification of affected parts of the \gls{ran} stack; for example, the presence of the strong MAC keyword \verb|MSG3| and \verb|MAC| yields a weighted score of 2.0, exceeding the MAC detection threshold and triggering layer identification.

Keyword-based categorization relies on lexical matching and does not capture semantic context, which can lead to misclassification. For example, while terms such as \verb|tx| and \verb|rx| are ambiguous in isolation, they can be disambiguated when combined with contextual keywords like \verb|L1| or \verb|MAC|. To address this limitation, commits that produce low or medium confidence scores are passed to an \gls{llm} for semantic refinement. We use Claude~4.5 Sonnet as the \gls{llm} for this stage.
\vspace{-1.5em}
\begin{figure}[h]
\centering
{\scriptsize
\setlength{\baselineskip}{0.9\baselineskip}
\begin{lstlisting}[frame=single, basicstyle=\ttfamily\tiny, breaklines=true, escapeinside={(*}{*)}]
2557: fix duplicate call of RCconfig_(*\textbf{NR}*)_(*\textbf{L1}*)
2556: Support RC SM aperiodic subscription for "UE (*\textbf{RRC}*) State Change"
2550: use pointer to structure instead of module_id inside (*\textbf{MAC}*)
2548: NR UE (*\textbf{MSG3 buffer}*)
2495: Sidelink configuration passed from (*\textbf{RRC}*)->(*\textbf{MAC}*)
2490: reworking configuration of (*\textbf{LogicalChannelConfig}*) at (*\textbf{MAC}*) UE
2220: (*\textbf{L1 tx thread}*)

2557 -> [PHY] ("L1","NR")
2556 -> [RRC] ("RRC")
2550 -> [MAC] ("MAC")
2548 -> [MAC] ("MAC","MSG3","buffer")
2495 -> [RRC,MAC] ("RRC","MAC")
2490 -> [MAC] ("MAC","LogicalChannelConfig")
2220 -> [PHY] ("L1","tx","thread")
\end{lstlisting}
}
\vspace{-0.2cm}
\caption{Keyword-based categorization examples.}
\label{fig:keyword_categorization}
\vspace{-0.3cm}
\end{figure}

\subsubsection{\gls{llm}-Based Refinement}
\label{sec:llm}

The \gls{llm} is provided with a fixed, human-designed instruction prompt together with the commit text and the results of the keyword-based classification. It is instructed to perform structured validation by (i) confirming or rejecting candidate layers and components to eliminate false positives from keyword matching, (ii) enforcing an upper bound of four affected layers to prevent over-classification, and (iii) identifying a single primary change type. The model returns a structured output containing the refined layer set, component set, primary change category, and a brief justification. The output of this refinement stage is the finalized set of affected protocol layers and functional components, $\Gamma_c$, along with an associated confidence indicator. Invoking the \gls{llm} only for commits with ambiguous or low-confidence keyword evidence preserves scalability while improving semantic accuracy.

The final vector $\mathbf{f}_c$ produced with 34 features for each commit includes binary layer and component indicators, normalized change-type scores, aggregate counts (number of affected layers and components), code-churn metrics, and a composite change-complexity score that summarizes the overall scope and structural impact of the modification.

\subsection{Variance Decomposition}
\label{sec:variance-decompose}

In the second step of \framework, we quantify dependencies of extracted features from Section~\ref{sec:kpi-kpm-features} alongside code change features from Section~\ref{sec:hybrid} using variance decomposition. We want to determine the relative contributions of environmental factors versus code modifications to performance variation. This analysis enables us to quantify the marginal effect of each feature on a given performance metric and to track how these contributions evolve over time across successive software updates and in relation to one another.

Variance decomposition quantifies how much additional variance in a target metric $Y$ is explained by a set of predictors $P$, while controlling for another set of variables $Q$. This approach follows principles commonly used in ANOVA and regression analysis, and answers the question: ``how much more of the performance variation can be explained by adding a specific factor, after accounting for other confounding variables?'' In our case, we separately evaluate the contribution of channel conditions, traffic load, and code changes while controlling for the remaining factors.

To enable statistical grouping and conditional expectation computation, continuous \gls{kpi}/\gls{kpm} metrics are discretized into ordered categories. The variance decomposition score is computed using Equation~\eqref{eq:variance-decompose}. By the law of total variance, $Var(E[Y|X])$ represents the portion of variance in $Y$ that can be explained by $X$. The proposed metric captures the additional variance explained by predictors 
$P$ beyond what is explained by the variables $Q$, normalized by the total variance of $Y$.
\begin{equation}
    C_{var}(Y, P | Q) = \frac{\mathrm{Var}(\mathbb{E}[Y \mid P, Q]) - \mathrm{Var}(\mathbb{E}[Y \mid Q])}{\mathrm{Var}(Y)}
    \label{eq:variance-decompose}
\end{equation}

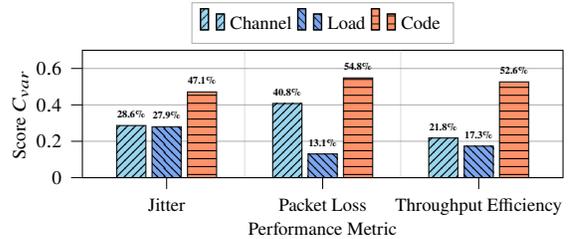
\begin{figure}[h]
  \centering
  \setlength\fwidth{0.9\linewidth}  
    \setlength\fheight{0.37\linewidth}
\tikzset{draw-color/.style={
        color of colormap={#1},
        draw=.!80,
    },
    fill-color/.style={
        color of colormap={#1},
        draw=.!80!black,
        fill=.!80!white,
        fill opacity=0.6
    },
    mydashed/.style={dash pattern=on 6pt off 4pt}
}

\definecolor{coral}{RGB}{255,127,80}
\definecolor{cornflowerblue}{RGB}{100,149,237}
\definecolor{darkgray176}{RGB}{176,176,176}
\definecolor{gray}{RGB}{128,128,128}
\definecolor{skyblue}{RGB}{135,206,235}

\begin{tikzpicture}
\begin{axis}[
width=\fwidth,
height=\fheight,
legend cell align={left},
legend style={
  font=\scriptsize,
  fill opacity=0.8,
  draw opacity=1,
  text opacity=1,
  at={(0.5,1.05)},
  anchor=south,
  draw=darkgray176,
  font=\scriptsize
},
tick align=outside,
tick pos=left,
x grid style={darkgray176},
xlabel={Performance Metric},
xlabel style={font=\scriptsize},
xmin=-0.54, xmax=2.54,
xtick style={color=black},
xtick={0,1,2},
xticklabels={Jitter,Packet Loss,Throughput Efficiency},
tick label style={font=\scriptsize},
y grid style={darkgray176},
ylabel={Score $C_{var}$},
ylabel style={font=\scriptsize},
ymajorgrids,
ymin=0, ymax=0.7,
ytick style={color=black},
xlabel shift=-4pt,
ylabel shift=-4pt,
legend columns=3,
font=\scriptsize
]
% Channel - North East Lines (///) with skyblue background
\draw[draw=black,fill=skyblue,opacity=0.85,line width=0.48pt] (axis cs:-0.32,0) rectangle (axis cs:-0.133333333333333,0.286291558605249);
\draw[draw=none,pattern=north east lines,pattern color=black,line width=0.48pt] (axis cs:-0.32,0) rectangle (axis cs:-0.133333333333333,0.286291558605249);
\addlegendimage{ybar,ybar legend,draw=black,fill=skyblue,opacity=0.85,postaction={pattern=north east lines,pattern color=black},line width=0.48pt}
\addlegendentry{Channel}

\draw[draw=black,fill=skyblue,opacity=0.85,line width=0.48pt] (axis cs:0.68,0) rectangle (axis cs:0.866666666666667,0.407503813120095);
\draw[draw=none,pattern=north east lines,pattern color=black,line width=0.48pt] (axis cs:0.68,0) rectangle (axis cs:0.866666666666667,0.407503813120095);

\draw[draw=black,fill=skyblue,opacity=0.85,line width=0.48pt] (axis cs:1.68,0) rectangle (axis cs:1.866666666666667,0.218224204882104);
\draw[draw=none,pattern=north east lines,pattern color=black,line width=0.48pt] (axis cs:1.68,0) rectangle (axis cs:1.866666666666667,0.218224204882104);

% Load - North West Lines (\\\) with cornflowerblue background
\draw[draw=black,fill=cornflowerblue,opacity=0.85,line width=0.48pt] (axis cs:-0.093333333333333,0) rectangle (axis cs:0.093333333333333,0.279471252009665);
\draw[draw=none,pattern=north west lines,pattern color=black,line width=0.48pt] (axis cs:-0.093333333333333,0) rectangle (axis cs:0.093333333333333,0.279471252009665);
\addlegendimage{ybar,ybar legend,draw=black,fill=cornflowerblue,opacity=0.85,postaction={pattern=north west lines,pattern color=black},line width=0.48pt}
\addlegendentry{Load}

\draw[draw=black,fill=cornflowerblue,opacity=0.85,line width=0.48pt] (axis cs:0.906666666666667,0) rectangle (axis cs:1.093333333333333,0.130518656266928);
\draw[draw=none,pattern=north west lines,pattern color=black,line width=0.48pt] (axis cs:0.906666666666667,0) rectangle (axis cs:1.093333333333333,0.130518656266928);

\draw[draw=black,fill=cornflowerblue,opacity=0.85,line width=0.48pt] (axis cs:1.906666666666667,0) rectangle (axis cs:2.093333333333333,0.173461999434163);
\draw[draw=none,pattern=north west lines,pattern color=black,line width=0.48pt] (axis cs:1.906666666666667,0) rectangle (axis cs:2.093333333333333,0.173461999434163);

% Code - Horizontal Lines with coral background
\draw[draw=black,fill=coral,opacity=0.85,line width=0.48pt] (axis cs:0.133333333333333,0) rectangle (axis cs:0.32,0.470901315485469);
\draw[draw=none,pattern=horizontal lines,pattern color=black,line width=0.48pt] (axis cs:0.133333333333333,0) rectangle (axis cs:0.32,0.470901315485469);
\addlegendimage{ybar,ybar legend,draw=black,fill=coral,opacity=0.85,postaction={pattern=horizontal lines,pattern color=black},line width=0.48pt}
\addlegendentry{Code}

\draw[draw=black,fill=coral,opacity=0.85,line width=0.48pt] (axis cs:1.133333333333333,0) rectangle (axis cs:1.32,0.547709069155597);
\draw[draw=none,pattern=horizontal lines,pattern color=black,line width=0.48pt] (axis cs:1.133333333333333,0) rectangle (axis cs:1.32,0.547709069155597);

\draw[draw=black,fill=coral,opacity=0.85,line width=0.48pt] (axis cs:2.133333333333333,0) rectangle (axis cs:2.32,0.525894595975456);
\draw[draw=none,pattern=horizontal lines,pattern color=black,line width=0.48pt] (axis cs:2.133333333333333,0) rectangle (axis cs:2.32,0.525894595975456);

% Separator lines
\addplot [very thin, gray, opacity=0.3, forget plot]
table {%
0.5 0
0.5 0.7
};
\addplot [very thin, gray, opacity=0.3, forget plot]
table {%
1.5 0
1.5 0.7
};

% Percentage labels
\draw (axis cs:-0.226666666666667,0.301291558605249) node[
  scale=0.55,
  anchor=south,
  text=black,
  rotate=0.0
]{\bfseries 28.6\%};
\draw (axis cs:0.773333333333333,0.422503813120095) node[
  scale=0.55,
  anchor=south,
  text=black,
  rotate=0.0
]{\bfseries 40.8\%};
\draw (axis cs:1.773333333333333,0.233224204882104) node[
  scale=0.55,
  anchor=south,
  text=black,
  rotate=0.0
]{\bfseries 21.8\%};
\draw (axis cs:0,0.294471252009665) node[
  scale=0.55,
  anchor=south,
  text=black,
  rotate=0.0
]{\bfseries 27.9\%};
\draw (axis cs:1,0.145518656266928) node[
  scale=0.55,
  anchor=south,
  text=black,
  rotate=0.0
]{\bfseries 13.1\%};
\draw (axis cs:2,0.188461999434163) node[
  scale=0.55,
  anchor=south,
  text=black,
  rotate=0.0
]{\bfseries 17.3\%};
\draw (axis cs:0.226666666666667,0.485901315485469) node[
  scale=0.55,
  anchor=south,
  text=black,
  rotate=0.0
]{\bfseries 47.1\%};
\draw (axis cs:1.226666666666667,0.562709069155597) node[
  scale=0.55,
  anchor=south,
  text=black,
  rotate=0.0
]{\bfseries 54.8\%};
\draw (axis cs:2.226666666666667,0.540894595975456) node[
  scale=0.55,
  anchor=south,
  text=black,
  rotate=0.0
]{\bfseries 52.6\%};
\end{axis}
\end{tikzpicture}  
\vspace{-.6em}
   \caption{Variance decomposition for affecting factors.}
    \label{fig:variance-decompose}
\end{figure}

Figure~\ref{fig:variance-decompose} presents variance decomposition results 
for throughput efficiency, packet loss, and jitter, showing the proportion 
of variance explained by each factor  while controlling for the others. As expected, channel-related features and traffic load explain a significant portion of the performance variation. The impact of channel conditions is most pronounced for packet loss, accounting for approximately $40\%$ of the variance, which is more than double its effect on throughput efficiency. In contrast, traffic load has a stronger influence on jitter, explaining about 27\% of its variance, compared to a much smaller contribution for packet loss. 
We also observe that code changes exhibit significant and consistent contribution to performance variation across all metrics. We proceed to residual-based analysis after confirming the contribution of code-change on performance. Periodic variance decomposition checks can track temporal shifts in the balance between environmental and code-induced effects as the code evolves.
\vspace{-.8em}

\subsection{Environment Model}
\label{sec:env-model}

To isolate code-induced performance variations, we first establish an environmental baseline predicting expected performance under given environmental conditions. 
This model effectively answers the question: ``what throughput efficiency should we ideally achieve given the network conditions and traffic load?'' 

To predict baseline throughput from environmental conditions, we evaluate several machine learning models including Random Forest, XGBoost, Decision Tree, Linear Regression, and K-Nearest Neighbors. The input features to the model include channel quality indicators like \gls{snr}, \gls{rsrp}, link layer performance 
metrics like HARQ retransmission statistics, traffic and load characteristics including target data rate. 

\begin{figure}[t]
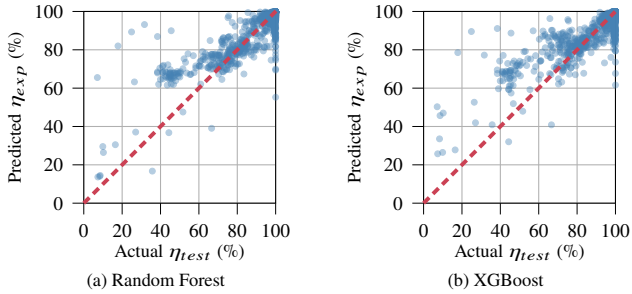

  \centering
  \begin{subfigure}[b]{0.48\columnwidth}
    \centering
    \setlength\fwidth{0.6\linewidth}  
    \setlength\fheight{0.6\linewidth}
    \input{figures/tikz-codes/random-forest-scatter_1}
    \vspace{-0.2cm}
    \caption{Random Forest}
    \label{fig:random-forest}
  \end{subfigure}
  \hfill
  \hspace{-1em}
  \begin{subfigure}[b]{0.48\columnwidth}
    \centering
    \setlength\fwidth{0.6\linewidth}  
    \setlength\fheight{0.6\linewidth}
    \input{figures/tikz-codes/xgboost-scatter_1}
      \vspace{-0.2cm}
    \caption{XGBoost}
    \label{fig:xgboost}
  \end{subfigure}
  \vspace{-.2em}
  \caption{Predicted vs. actual throughput efficiency for the best learning models.}
  \label{fig:model-scatter-all}
\end{figure}

We show the results of two best environment models and their quantitative performance metrics in
Figure~\ref{fig:model-scatter-all} and Table~\ref{tab:model_comparison}. The figure reports how much the predicted values (y axis) deviate from the actual values (x axis). Random Forest achieves the best overall performance with $R^2 = 0.767$, 
\gls{rmse} = 5.626 Mbps, and \gls{mae} = 2.17 Mbps. We configure the Random Forest regressor with an ensemble of 100 decision trees, with maximum depth of 8 levels to balance expressiveness with generalization.

\begin{table}[h]
\centering
\caption{Environmental model performance for the best learning models.}
\label{tab:model_comparison}
\begin{tabular}{lcccc}
\toprule
\textbf{Model} & \textbf{Test $\mathbf{R^2}$} & \textbf{ MAE } & \textbf{ RMSE }\\
\midrule
Random Forest         & 0.767 & 2.170 & 5.623 \\
XGBoost              & 0.725 & 2.418 & 6.105 \\
\bottomrule
\end{tabular}
\end{table}
\vspace{-1.2em}

\subsection{Residual-Based Performance Analysis}
Once we have the baseline estimation of the throughput performance that can be obtained based on the current conditions, we can isolate the anomalies in performance by calculating residuals. Formally, the residual of a test instance is defined as the performance ratio:
\begin{equation}
    \rho = \eta_{test}/\eta_{exp}
\end{equation}
where $\eta_{test}$ is the throughput efficiency observed in the test instance and $\eta_{exp}$ is the expected throughput efficiency given the environment. With residual values, we can label the observed performance as an anomaly or not. 
A test instance is labeled as degraded 
if two conditions are met: (i)~the residual falls below a degradation threshold 
$\tau_\rho$ ($\rho < \tau_\rho$), indicating under-performance, while (ii)~the expected throughput exceeds a minimum threshold $\tau_{exp}$ ($\eta_{exp} \geq \tau_{exp}$), confirming that environment conditions were sufficiently favorable to support normal operations. The second criterion is essential because wireless networks exhibit inherent variability. In conditions where $\eta_{exp} < \tau_{exp}$, environmental factors dominate performance variation, making it unreliable to attribute degradations to code changes. We select $\tau_{exp}$ by analyzing the trade-off between test coverage and detection reliability, ensuring we capture degradations in realistic operating conditions. 
For each instance labeled as degraded, we attribute the performance deviation 
to specific protocol layers by mapping the residual value $\rho$ with the set of modified layers, establishing a direct link between code changes and performance impact.

\subsection{Performance Risk Estimation}
Having labeled training instances based on residual analysis, we train a classifier to predict the probability of code-induced degradation for new commits. This can enable proactive risk assessment in continuous integration workflows. 

We leverage a gradient boosting classifier (LightGBM) that uses both environmental features (SNR, BLER, target rate) and code changes (affected categories $\mathbf{f}_c$) as features. The classifier is trained with 400 estimators, maximum depth 4, and learning rate 0.1 to predict performance degradation.  
To address class imbalance in our performance dataset, we employ \gls{smote} \cite{chawla2002smote}. This generates synthetic samples for the degraded performance class, which is the minority to balance the class distribution. \gls{smote} creates new minority class instances by interpolating between existing samples in the feature space, which helps prevent overfitting. We also employ balanced class weighting to assign higher penalties to the minority class during training. The model outputs a degradation probability score for each test instance, enabling proactive identification of high-risk commits prior to deployment.

\section{Evaluation and Results}
\label{sec:evaluation}

In this section, we discuss \framework results related to \gls{llm}-refinement on a commit, degradation classification, the analysis on the code change impact, and the results on performance risk estimation.
\begin{figure}[b]
\vspace{-1em}
\centering
{\scriptsize
\setlength{\baselineskip}{0.9\baselineskip}
\begin{lstlisting}[frame=single, basicstyle=\ttfamily\tiny, breaklines=true, escapeinside={(*}{*)}]
* !2680 CI: Modification of log collection in UndeployObject()
* !2681 remove a useless copy and specific (*\textbf{buffer}*) for all UE UL payload
* !2685 Clang: make executable run, fix clang warnings, fix memsan warnings
* !2690 Remove hardcoding of 5G-S-TMSI on nrUE
\end{lstlisting}
}
\vspace{-.6em}
\caption{Commit example which triggered \gls{llm} refinement.}
\label{fig:llm-refinement}
\end{figure}

\subsection{\gls{llm}-Refinement}
\label{sec:llm-refinement}
This section illustrates the necessity of \gls{llm}-based refinement using a representative example. Figure~\ref{fig:llm-refinement} shows commit 2024.w16, where the keyword-based categorization stage failed to identify any protocol layers due to insufficient keywords, detecting only the memory component based on the term \textit{buffer} and resulting in a low confidence score. The \gls{llm} refinement was therefore triggered and was able to infer the affected layers by interpreting the semantic context of the commit messages, associating ``UE UL payload'' modifications with the MAC layer and the ``remove hardcoding of 5G-S-TMSI on nrUE'' with the NAS layer.

\subsection{Residual Analysis and Degradation Classification}
\label{sec:residual-analysis-degradation-classification}

\begin{figure}[h]
  \centering
  \includegraphics[width=0.9\columnwidth]{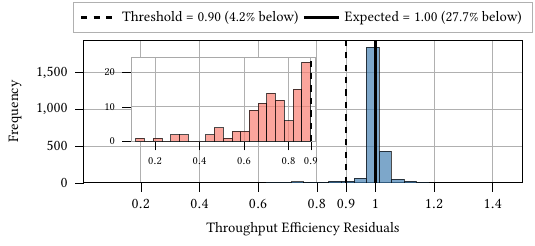}
  \vspace{-.6em}
   \caption{Throughput efficiency residuals distribution.}
    \label{fig:residuals}
    \vspace{-.3cm}
\end{figure}

We first examine the residual distribution to validate our environmental model. Figure~\ref{fig:residuals} shows the distribution of residuals $\rho$ across all test instances. The majority of residuals 
cluster around $\rho \approx 1.0$ with mean = $0.998$ and 
median = $1.00$, indicating 
performance aligned with environmental expectations. We choose $\tau_\rho = 0.9$ as threshold for under-performance and  4.2\% of instances exhibit $\rho < 0.9$. A small Welch's $t$-test $p$-value ($<0.001$) and Cohen's $d$ of $2.45$ suggest that the two groups are distinct and validate the choice of the $0.9$ threshold. The degraded group, with mean around $0.7$ and standard deviation $0.159$, has a three times higher variability than the group of residuals above the threshold with standard deviation $0.053$.
\vspace{-1em}

Next, we select the expected throughput threshold $\tau_{exp}$ that determines which test instances have adequate environmental conditions for reliable code attribution. Higher thresholds (e.g., 80\%) would miss degradations in moderate channel conditions, while lower ones (e.g., 40\%) would produce false positives from environmental variability. We analyze the trade-off between test coverage and detection reliability across thresholds from 40\% to 90\%. Based on this analysis, we select $\tau_{exp} = 60\%$, which provides optimal balance, retaining 99.5\% of test instances with standard deviation of $0.068$ enabling reliable analysis. 

\begin{figure}[h]
    \centering
\includegraphics[width=0.85\columnwidth]{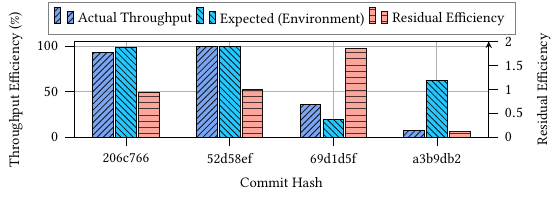}
\vspace{-.6em}
    \caption{Example cases.}
    \label{fig:commit_residual}
\end{figure}

To demonstrate how the residual-based framework distinguishes code-based degradations from environmental effects, in Figure~\ref{fig:commit_residual}, we present three representative evaluation cases.

\textbf{Case 1 - Normal operation:} In this case, the test achieves similar performance as the baseline prediction. The first two commits of Figure~ \ref{fig:commit_residual} are examples of this case, where the
residual $\rho$ is $1$ and $0.95$, respectively, and thus falls within normal environmental fluctuation. This is treated as normal operation.

\textbf{Case 2 - Environmental limitation:} In this case, baseline and test throughput efficiency results are much lower than 60\%, indicating degraded environmental conditions. Under 
such poor conditions, $\rho$ loses reliability and we cannot confidently correlate under-performance to code changes because the environment itself fundamentally limits the achievable throughput efficiency. Thus, our framework prevents degradation flagging in these cases.
This case is observed for commit \texttt{69d1d5f}. We also observe that when the commit is tested under normal environmental conditions, no performance degradation occurs.

\textbf{Case 3 - Code-induced degradation:} In this case, the baseline throughput efficiency is higher but the test result is much lower. A low residual value $\rho$, combined with favorable baseline, triggers degradation flagging. This is observed for commit \texttt{a3b9db2}, which
upgraded NVIDIA Aerial (PHY acceleration) from version 24-3 to 25-1. 
A commit branching from \texttt{a3b9db2} fixed PHY bugs, where the SINR was wrongly set to 0 for SISO and the CQI bit length was incorrectly calculated. 
These bugs were likely regressions from the Aerial 25-1 version upgrade.

\begin{figure}[h]
    \centering
\input{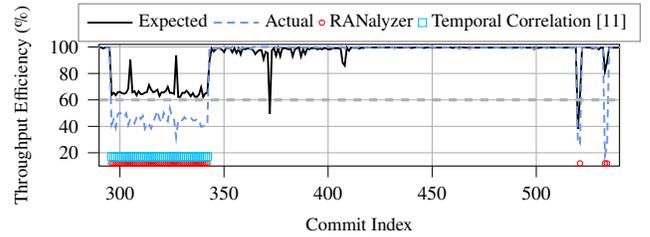}
    \vspace{-.6em}
    \caption{Baseline comparison.}
    \label{fig:baseline}
    \vspace{-1em}
\end{figure}

We compare our residual-based degradation detection against a temporal correlation baseline adapted from~\cite{li2020gandalf}. This approach uses multi-window aggregation with exponential weights to prioritize faults 
occurring shortly after code deployment, based on the intuition that temporal 
proximity indicates causal relationships. 
Figure~\ref{fig:baseline} shows flagging patterns using this approach across a sample set of test instances for target throughput rate of $30$\:Mbps. Our proposed method detects degradation when there is a series of degradation with expected throughput ${>}\,60\%$. 
   \vspace{-1em}
The temporal-correlation-based method, however, misses certain degradations that happens with a longer delay after the code change, as their temporal distance reduces their contribution under the exponential weighting.
Our residual analysis explicitly models environmental baselines, achieving better precision by only flagging under-performance when conditions were favorable.

\subsection{Residual-Based Analysis of Code Change Impact}
We analyze the relationship between code changes and performance degradation leveraging residuals derived from the environment model. Table~\ref{tab:layer_degradation} presents performance degradation statistics for each protocol layer, showing the distribution of residual values for commits modifying each layer.

We observe that PDCP modifications impact degradation the most, with mean residual of $0.70$, followed by RRC and NGAP. PDCP shows particularly high variance (standard deviation of $0.28$), indicating inconsistent impact on performance. The large median-mean gap for this layer shows that the distribution is skewed by a subset of commits with extreme degradation. Meanwhile, we note that MAC changes have the highest number of degraded cases. One reason may be the high frequency of MAC modifications in the development process. NAS modifications, instead, show less severe degradation with lower variation.
\begin{table}[h]
\centering
\caption{Performance impact by modified protocol layer.}
\label{tab:layer_degradation}
\begin{tabular}{lcccc}
\toprule
\textbf{Layer} & \textbf{Degraded Cases} & \textbf{Mean $\rho$} & \textbf{Median $\rho$} & \textbf{Std Dev} \\
\midrule
PDCP & 14 & 0.70 & 0.83 & 0.28 \\
RRC & 87 & 0.72 & 0.73 & 0.15 \\
NGAP & 87 & 0.72 & 0.73 & 0.15 \\
MAC & 101 & 0.74 & 0.75 & 0.15 \\
PHY & 87 & 0.74 & 0.74 & 0.12 \\
NAS & 40 & 0.78 & 0.84 & 0.19 \\
\bottomrule
\end{tabular}
\end{table}
\vspace{-.6em}

This layer-specific analysis enables risk-based code reviews. For example, commits modifying PDCP, RRC or NGAP may require enhanced scrutiny due to their larger average impact, while MAC changes require attention primarily due to high frequency despite average degradation. PHY changes, despite moderate severity, are more predictable and may be easier to validate through targeted testing.

\subsection{Results on Performance Risk Estimation}
We now evaluate the result of the performance risk estimation. The classifier that predicts code-induced degradation achieves excellent performance on majority class with precision = $0.98$, recall = $0.94$, F1 = $0.96$. This indicates reliable identification of commits that do not degrade performance. With only 18 instances in the test set, the classifier achieves recall = $0.61$ (detecting 11 out of 18 true degradations) but precision = $0.26$ (26\% of degradation predictions are correct). 
This is expected given the challenges of severe class imbalance, high variability in performance outcomes, and prediction from commit metadata and environmental context without using actual test performance measurements.
The low precision indicates false alarms, classifying benign commits as degrading, though the moderate recall demonstrates the ability to identify genuine degradation cases. The recall-precision trade-off is acceptable for \gls{ci}/\gls{cd} where missing a degradation 
(false negative) is more costly than investigating a false alarm.

\section{Conclusions}
\label{sec:conclusion}
In this paper, we proposed \framework to address the challenge of attributing performance variations in software-defined cellular networks. Here, observed metrics reflect the joint impact of software changes and environmental conditions. Our approach combined (i) hybrid software release categorization, using keyword pattern matching with selective LLM refinement to construct a code-change feature vector, with (ii) environment-controlled residual analysis, which models baseline performance from channel and traffic conditions and isolates code-induced performance deviations.

Evaluation on more than two years of \gls{oai} 5G \gls{ci}/\gls{cd}/\gls{ct} data comprising 69 software releases and more than 8600 tests demonstrated the \framework effectiveness. Our approach identified code-induced degradations, distinguishing between tests that underperform due to channel conditions or software regression. Case studies validated practical utility across bug fixes, 
optimizations, and refactoring, with layer-specific attribution accelerating 
root cause identification. In future extensions, \framework can evolve through continuous data accumulation enabling model refinement, fine-grained event metrics extraction, and multi-level severity classification for nuanced 
risk assessment.

\bibliographystyle{IEEEtran}
\bibliography{biblio}

\end{document}